\documentstyle[psfig,epsf,amssymb]{elsart}
\begin{document}
\begin{frontmatter}
\title{Optimal local discrimination of two multipartite pure states}
\author{S. Virmani$^{1}$, M. F. Sacchi$^{1,2}$, M. B. Plenio$^1$ and D. Markham$^1$}
\address{$^1$ Optics Section, Blackett Laboratory, Imperial College, 
London SW7 2BW, United Kingdom.}
\address{$^2$ Dipartimento di Fisica `A. Volta', Universit\`{a} di Pavia and Unit\`{a} INFM, via A. Bassi 6, I27100 Pavia, Italy.}
\date{\today}\maketitle
\begin{abstract}
In a recent paper, Walgate et. al. \cite{Wal SHV} demonstrated that any two
{\it orthogonal} multipartite pure states can be optimally distinguished using
only local operations. We utilise their result to show that this is
true for {\it any} two multiparty pure states, in the sense of inconclusive
discrimination. There are also certain regimes of conclusive discrimination for
which the same also applies, although we can only conjecture that the result is
true for all conclusive regimes. We also discuss a class of states that can be distinguished locally according to {\it any} discrimination measure, as they can be locally recreated in the possession of one party. A consequence of this is that any two maximally entangled states can always be optimally discriminated locally, according to {\it any} figure of merit. 
\end{abstract}
\end{frontmatter}
\section{Introduction}

Entanglement lies at the heart of many aspects of
quantum information theory and it is therefore desirable to understand its
structure as well as possible. One attempt to improve our understanding of
entanglement is the study of our ability to perform information theoretic tasks
locally on non-local states, such as the local implementation of
non-local quantum gates \cite{gates}, telecloning \cite{Murao 99}, the remote
manipulation and preparation of quantum states \cite{remote} or the recently
studied question of the local discrimination of non-local states by a variety
of authors. In \cite{Wal SHV} it was shown that any two orthogonal pure states
can be perfectly discriminated locally, whereas in \cite{Ter DL} examples of
two {\it orthogonal} mixed states were presented which {\it cannot} be
distinguished perfectly locally. Another surprising development is that there
exist bases of product orthogonal pure states which cannot be locally reliably
discriminated, despite the fact that each state in the basis contains no
entanglement \cite{Ben DF+}. Here we discuss the issue of discriminating two
non-orthogonal pure states locally, and show that in this regime the optimal
global procedure can be achieved.

\section{Background Material}

We will begin by recalling the optimal method for discriminating two states
when we are {\it not} restricted to local actions. Suppose that we are given
one of two mixed states $\rho_{1}$ and $\rho_{2}$ (not necessarily orthogonal)
with prior probabilities $p_{1}$ and $p_{2}$ respectively, and we have been
asked to optimally determine which state we have actually been
given. Helstrom \cite{Helstrom} (see also \cite{Fuchsthesis} for a 
very readable account) 
showed that the optimal discrimination protocol is as
follows. Consider the following Hermitian matrix:
\begin{equation}
    \Delta = p_{1}\rho_{1} -  p_{2}\rho_{2}
\end{equation}
As it is Hermitian, $\Delta$ has a basis of orthogonal
eigenvectors $\{|\delta_{i}\rangle\}$. In general $k$ of these
eigenstates will have positive (or zero) eigenvalues and $n-k$
will be negative, where $n$ is the dimension of $\Delta$. We will
denote the positive eigenstates as $\{|\delta_{i}^{+}\rangle\},
i=1...k $ and the negative eigenstates as
$\{|\delta_{i}^{-}\rangle\}, i=k+1...n$. The optimal measurement
is then to measure in this basis and whenever one of the positive
eigenstates is obtained return the answer $\rho_{1}$ and whenever
a negative eigenstate is obtained return the answer $\rho_{2}$. In
this case the probability of error will be given by:

\begin{equation}
    P_{E} =\sum_{i=1}^{k} p_{2} \hbox{tr} ( \rho_{2} | \delta_{i}^{+} \rangle
    \langle \delta_{i}^{+}|) + \sum_{i=k+1}^{n} p_{1}\hbox{tr} ( \rho_{1} |
    \delta_{i}^{-} \rangle \langle \delta_{i}^{-}|)
\label{error}
\end{equation}

This is the optimal probability which can be achieved by any POVM
\cite{Helstrom}. In particular, in the case of two pure states
$|\psi_{1}\rangle,|\psi_{2}\rangle$ with prior probabilities $p_{1}$ and
$p_{2}$ respectively, the optimal measurement consists of measuring in the
basis $\{|+\rangle,|-\rangle\}$, where  $\{|+\rangle,|-\rangle\}$ are the
positive and negative eigenvalue eigenstates of the matrix $\Delta =
p_{1}|\psi_{1}\rangle\langle\psi_{1}| - p_{2}|\psi_{2}\rangle\langle\psi_{2}|$.
Although in this case $\Delta$ will in general also have many eigenstates of
eigenvalue zero, they are not important as they will never occur in any
measurement of the two pure states. The probability of error is then given by:
\begin{equation}
    P_{E} = p_{2} \hbox{tr} (  | \psi_{2} \rangle \langle \psi_{2} | + \rangle \langle +|) +
     p_{1} \hbox{tr} (  | \psi_{1} \rangle \langle \psi_{1} | - \rangle \langle -|)
\end{equation}
It will be useful to note that the states $\{|+\rangle,|-\rangle\}$ are constructed in such a way that we can always write:
\begin{eqnarray}
\label{useful}
    |\psi_{1}\rangle = a|+\rangle + b|-\rangle \nonumber\\
    |\psi_{2}\rangle = c|+\rangle + d|-\rangle
\end{eqnarray}
for some complex coefficients $a,b,c,d$.

We will later require the results of Walgate et. al. \cite{Wal SHV} to show that this
probability of error can be achieved using a locally implementable measurement.
They showed that any two bipartite orthogonal pure states $|+\rangle$ and
$|-\rangle$ on Hilbert space $H_{A} \otimes H_{B}$ can be expressed in the
following way:
\begin{eqnarray}
    |+\rangle = \sum_{i} \alpha_{i}|i\rangle|\eta_{i}\rangle \nonumber \\
    |-\rangle = \sum_{i} \beta_{i}|i\rangle|\mu_{i}\rangle
\label{wdec}
\end{eqnarray}
where the $\{|i\rangle\}$ form an orthonormal basis for $H_{A}$ and the
$|\eta_{i}\rangle,|\mu_{i}\rangle$ are normalised states satisfying:
\begin{equation}
    \langle\mu_{i}|\eta_{i}\rangle = 0.
\end{equation}
Other than this condition the $|\eta_{i}\rangle$ and
$|\mu _{i}\rangle$ needn't be orthogonal, although  we will treat them as
normalised (this is slightly different to the convention used in
\cite{Wal SHV}). By proving the existence of such a decomposition, \cite{Wal
SHV} showed how two orthogonal pure states can be discriminated locally. First
Alice must measure her side in the basis $\{|i\rangle\}$. If she obtains the
outcome $|j\rangle$, Bob must then measure in an orthonormal basis containing
the states $|\eta_{j}\rangle$ and $|\mu_{j}\rangle$. As
$\langle\mu _{i}|\eta_{i}\rangle = 0~~\forall i$, this is always possible. If
the outcome corresponds to $|\eta_{j}\rangle$ then the state was $|+\rangle$,
if the outcome corresponds to $|\mu_{j}\rangle$ then the state was $|-\rangle$.
In more formal terms this corresponds to performing a locally-implementable
measurement with elements $\{\{|i\eta_{i}\rangle\langle
i\eta_{i}|\},\{|i\mu_{i}\rangle\langle i\mu_{i}|\} \}$, such that if the
outcomes $\{|i\eta_{i}\rangle\langle i\eta_{i}|\}$ are obtained we know that
the state was $|+\rangle$ and if the outcomes $\{|i\mu_{i}\rangle\langle
i\mu_{i}|\}$ are obtained then the state was $|-\rangle$.

\section{Optimal local inconclusive discrimination of two multipartite pure states }

They key point in generalising this result to non-orthogonal states is that in
order to discriminate two nonorthogonal pure states $|\psi_{1}\rangle$ and
$|\psi_{2}\rangle$ optimally we need to measure in a basis of the two {\it
orthogonal} states $\{|+\rangle,|-\rangle\}$ given by the eigenvectors of
$\Delta$. So we can `pretend' that we are essentially trying to discriminate
between one of the two orthogonal states $\{|+\rangle,|-\rangle\}$, and
implement the local measurment $\{\{|i\eta_{i}\rangle\langle
i\eta_{i}|\},\{|i\mu_{i}\rangle\langle i\mu_{i}|\} \}$. If the outcomes
$\{|i\eta_{i}\rangle\langle i\eta_{i}|\}$ are obtained we say that the original
state was $|\psi_{1}\rangle$, whereas if the outcomes
$\{|i\mu_{i}\rangle\langle i\mu_{i}|\}$ are obtained we say that the original
state was $|\psi_{2}\rangle$. Indeed, with this POVM the probability of error
will be given by:

\begin{equation}
    P_{E} =\sum_{i} p_{2} \hbox{tr} ( |\psi_{2}\rangle\langle\psi_{2}|i\eta_{i}\rangle
    \langle i\eta_{i}|) +  p_{1}\hbox{tr} ( |\psi_{1}\rangle\langle\psi_{1}|i\mu_{i}\rangle
    \langle i\mu_{i}|)
\end{equation}

It is easy to verify (using equation (\ref{useful})) that this probability of
error is equal to that of equation (\ref{error}), thus demonstrating that we
can achieve the optimal discrimination probability for two bipartite pure
states by using only local operations.

The generalisation to multipartite pure states follows in a
similar manner. Walgate et. al. \cite{Wal SHV} pointed out that
the decomposition (\ref{wdec}) also implies that any two
multipartite orthogonal states can be discriminated using only
local actions. Imagine two orthogonal three party pure states
shared between Alice, Bob and Chris. Bob and Chris can at first be
grouped as one party, and then the decomposition (\ref{wdec}) can
be applied to the Alice-(Bob,Chris) split. Then Alice measures her
side according to the above bipartite protocol, leaving
(Bob-Chris) in one of two orthogonal states depending upon which
tripartite pure state they all originally shared. Then Bob \&
Chris can simply apply the bipartite protocol to tell which state
they had. The same method can easily be seen to generalize to any
two multipartite orthogonal pure states. This result in turn can
be used to show, following exactly the same argument as the above
bipartite one, that any two multiparty pure states can be
distinguished in an optimal manner using local operations only.

It is worth pointing out that the above argument also extends to
some mixed state scenarios. In particular, when the two mixed
states to be distinguished collectively span only a two
dimensional space then there will be only two non-trivial
eigenstates of the matrix $\Delta$. In this situation the above
reasoning implies that two mixed states with this property can be
optimally distinguished locally.

\section{Conclusive discrimination}

The above arguments apply to the case of {\it inconclusive}
discrimination, in which we have to say that the state we are
given is either $\rho_{1}$ or $\rho_{2}$, but we are allowed to
sometimes make errors, the probability of which me must minimise.
However, there is also the case of {\it conclusive}
discrimination, in which we are allowed to give three outcomes;
$\rho_{1}$, $\rho_{2}$ or `don't know'. In this case, we must
never be wrong when we return outcomes $\rho_{1}$ or $\rho_{2}$,
and we have to try to minimise the probability of getting the
`don't know' outcome. It is therefore also of interest to know how
to optimise conclusive local discrimination. The optimal
conclusive protocol for globally discriminating two pure states is
discussed in references \cite{idp-js,js}. In some situations this
protocol reduces to performing an orthogonal measurement of two
orthogonal states, in which case exactly the same arguments as
above apply again, and the optimal discrimination protocol can be
matched by a local one. In particular, this applies to situations
in which \cite{js} (taking $p_{2}\geq p_{1}$ ):
\begin{equation}
|\langle \psi_{1}| \psi_{2}\rangle| \geq ({p_{1} \over
p_{2}})^{1/2}.
\end{equation}
In which case the optimal discrimination probability is given by:
\begin{equation}
P_{success}=p_{1}(1-|\langle \psi_{1}| \psi_{2}\rangle|^{2}).
\end{equation}
However, the general optimal conclusive case requires three
outcomes to have non-zero probability (corresponding to
$\rho_{1}$, $\rho_{2}$ and `don't know'), thus requiring a minimum
of three orthogonal pure states to be distinguished locally. As
the question of locally discriminating three orthogonal pure
states is currently unanswered, we were not able to extend these
results to the entire conclusive case. However, we were able to
make some progress in some situations with equal prior
probabilities, $p_{1}=p_{2}$. In this situation, \cite{idp-js}
show that the optimal global discrimination probability for two
pure states is given by:
\begin{equation}
1-P_{E} = 1 - |\langle \psi_{1}|\psi_{2}\rangle |
\end{equation}
It is easy to verify that in the case of two product pure states
if Alice and Bob implement such a measurement independently, and
then compare their results, then the probability that they both
fail is equal to the minimal global failure probability. Hence two
product pure states of prior probability 1/2 can be locally
discriminated optimally. It has also recently been shown that any two product 
pure states with arbritrary prior probabilities can be optimally conclusively discriminated locally \cite{Chen}. For more general two qubit pure states,
the situation is not so straightforward. We considered trying to
perform local one-dimensional projections on Alice side which
would give her no information, and leave Bob's particle in
residual states which could perhaps be easily distinguished from
one another. Although this is a very restricted protocol, it is
more suitable for analysis than a more general scheme. We will
write the two pure states to be distinguished in terms of two
matrices $\Psi_{1}$ and $\Psi_{2}$ in the following manner (see, for
example \cite{dar}):
\begin{eqnarray}
&&|\psi_{1} \rangle =\sum_{i,j}(\Psi_{1})_{ij}|i \rangle |j
\rangle \;,\nonumber
\\& & |\psi_{2}\rangle =\sum_{i,j}(\Psi_{2})_{ij}|i \rangle |j \rangle.
\;,\label{matrix}
\end{eqnarray}
In this notation, if Alice performs a local measurement an orthonormal basis
consisting of the states $|\chi \rangle $ and $|\chi ^\bot \rangle $, then the
condition
\begin{eqnarray}
\langle \chi | \Psi_{1} \Psi_{1}^\dag |\chi \rangle = \langle \chi
| \Psi_{2} \Psi_{2}^\dag |\chi \rangle \;\label{cond}
\end{eqnarray}
guarantees that the states after Alice's measurement on Bob's side
are given with the same probability as the initial two pure states
$|\psi_{1}\rangle $ or $|\psi_{2}\rangle $. If Bob then were to
apply the optimal conclusive discrimination on his side, the
overall probability of successful discrimination would be given by
\begin{eqnarray}
\tilde P_D=1-|\langle \chi |\Psi_{2}\Psi_{1}^\dag |\chi \rangle |
-|\langle \chi ^\bot |\Psi_{2}\Psi_{1}^\dag |\chi ^\bot \rangle |
\;,\label{one}
\end{eqnarray}
whereas the global optimal probability is equal to
\begin{eqnarray}
P_D&=&1-|\langle \chi |\Psi_{2}\Psi_{1}^\dag |\chi \rangle +
\langle \chi ^\bot |\Psi_{2}\Psi_{1}^\dag |\chi ^\bot \rangle
|\nonumber \\& =& 1- \hbox{Trace}[\Psi_{2}\Psi_{1}^\dag
]=1-|\langle \psi_{1}|\psi_{2} \rangle | \;.\label{two}
\end{eqnarray}
If both equation (\ref{cond}) and $\tilde P_D=P_D$ are satisfied
then we can say that such a protocol is optimal. We have
numerically and analytically confirmed that this is the case for a
large set of states, although we were not able to prove that such 
a solution is always possible. In cases with unequal prior
probability this method does not always work optimally, although
there may still be room to improve the protocol by dropping the
condition that Alice learns nothing from her measurement. However,
in such situations a much more intractable system of equations and
inequalities has to be satisfied (see \cite{js}), 
and we were not able to obtain
any significant solutions.

\section{Local discrimination according to {\it any} figure of merit}

\par Given that we have considered the problem of
optimal distinguishability under two quality measures, one could
ask the question as to whether there are states that can be
distinguished locally as well as globally according to {\it any}
discrimination figure of merit. An interesting example of states
for which this is possible is the set of {\em Schmidt correlated
states} \cite{Rains,note}. The method for distinguishing these
states is in fact very powerful, as it leads to the solution of a
more general problem. The reason for this is that we can transform
these states locally into local states with the same
distinguishability properties. Consider the special case of two
Schmidt correlated pure states, say:
\begin{eqnarray}
    \phi_{1} = \alpha|00\rangle + \beta |11\rangle \nonumber\\
    \phi_{2} = \gamma|00\rangle + \delta |11\rangle.
\label{maxcor}
\end{eqnarray}
Imagine that we have been given one of these states with some
arbitrary prior probabilities. If Alice performs a measurement in
the basis $\{|0\rangle+|1\rangle, |0\rangle-|1\rangle\}$, then she
will leave Bob's particle in some residual state. It is easy to check
that if Alice informs
Bob of the measurement outcome, then he can easily rotate his
state back to either $\alpha|0\rangle+\beta|1\rangle$ or
$\gamma|0\rangle+\delta|1\rangle$, depending upon whether
$\phi_{1}$ or $\phi_{2}$ respectively was originally shared (this solution 
can also be found according to the discussion
in Eqs. (\ref{cond}-\ref{two})). Bob
can then locally recreate the state that he and Alice shared by
adding an ancilla qubit in the state $|0\rangle$ and performing a
control-NOT operation. Surprisingly, this means that for Schmidt correlated
states the optimal global discrimination protocol can be achieved
locally, depsite the fact that the reduced density 
matrices are so similar. This method can easily be generalised to any number of Schmidt
correlated mixed states of any dimension \cite{Gen} and any number of
particles (where for more than two particles {\it Schmidt
correlated} would refer to mixtures of cat-states with the same cat-state basis).

The question then arises, when is it possible to express two pure states
in a Schmidt correlated form? In the following section we discuss 
necessary and sufficient conditions for this to be possible. These conditions are satisfied by many states, particularly by those with sufficient degeneracy amongst their Schmidt coefficients. An interesting consequence is that any two maximally entangled states can always be expressed in Schmidt correlated form, thus showing that two maximally entangled states can always be discriminated locally optimally, regardless of which figures of merit are chosen.

\section{When two pure states can be written in Schmidt Correlated form}

In this section we will discuss when it is possible to express two general pure
states in Schmidt correlated form. Using the matrix notation of equation (\ref{matrix}), it 
can be shown that a local basis change corresponding to unitaries $U \otimes V$ transforms 
the matrices representing the two pure states $|\psi_{1}\rangle$ and $|\psi_{2}\rangle$ as 
$U\Psi_{1}V^{T}$ and $U\Psi_{2}V^{T}$ respectively. Two states can be brought into maximally 
correlated form by such local unitaries if both the resulting state matrices are diagonal, i.e.:
\begin{eqnarray}
(U\Psi_{1}V^{T})_{ij} = x_{i} \delta_{ij} \nonumber \\
(U\Psi_{2}V^{T})_{ij} = y_{i} \delta_{ij}
\end{eqnarray}
for some complex $\{ x_{i}\}$ and $\{ y_{i}\}$. It is a well known consequence of the {\it 
singular value decomposition theorem} \cite{bhatia} of matrix analysis that any square 
matrix $\Psi$ can be brought into diagonal form by two unitaries $U,V$ acting as  $U\Psi V^{T}$.
Our question now concerns a modification of this result - when can two matrices $\Psi_{1}$ and 
$\Psi_{2}$ be brought into diagonal form by the {\it same} unitaries $U$ and $V$? In cases where 
the Schmidt decomposition of each pure state is unique, it may be easy to see if the 
two pure states are already Schmidt correlated. However, in cases where there is degeneracy amongst 
the Schmidt coefficients it is not always apparent whether some local basis change would bring both 
states into Schmidt correlated form. A necessary and sufficient condition is presented in \cite{Horn}:

{\bf Lemma 1:} Two square matrices $F$ and $G$ can be brought into diagonal form by the same unitaries $U,V$
iff {\it both} $FG^{\dag}$ and $G^{\dag}F$ are normal, i.e. they commute with their Hermitian conjugates.

In more physical terms, this means that two pure states can be brought into Schmidt correlated form by 
the same local unitaries iff the corresponding matrix representations $F$ and $G$ satisfy:
\begin{equation}
[FG^{\dag}, GF^{\dag}]=0,
\label{normal1}
\end{equation}
and
\begin{equation}
[G^{\dag}F, F^{\dag}G]=0.
\label{normal2}
\end{equation}

Although these conditions at first appear quite opaque, there are some cases where they give valuable insights. As mentioned earlier, an immediate consequence is that two maximally entangled states can always be brought into Schmidt correlated form. This surprising result follows from the fact that the reduced density matrices for a pure state represented by a matrix $F$ are given by $\rho_{A}(F)=FF^{\dag}$ and $\rho_{B}^{T}(F)=F^{\dag}F$. As maximally entangled states have reduced density matrices proportional to the identity, we can immediately see that any $F,G$ corresponding to maximally entangled states are proportional to unitary matrices. It is straightforward to verify that any such matrices satisfy the conditions of Lemma 1. In fact, we can easily go further and explicitly construct the local unitaries which will bring any two maximally entangled states into Schmidt correlated form. Without loss of generality, we can set the local bases such that $F$ is real and proportional to the identity (i.e. $F_{ij}=1/\sqrt{d}~\delta_{ij}$). Hence we can consider $F$ to be invariant under any unitary matrix $U$ acting as  $UFU^{\dag}$. As $G$ is unitary, it can be diagonalised by a unitary transformation $\sigma G \sigma^{\dag}$, where $\sigma$ is another unitary matrix. This corresponds to the physical operation $\sigma \otimes \sigma^{*}$. Hence $G$ will be made diagonal by such a transformation, and as $F$ is invariant under this operation, we will have succeeded in bringing both $F$ and $G$ into Schmidt correlated form.

It is in fact the degeneracy of the Schmidt coefficients of maximally entangled states which allows them to be brought into Schmidt correlated form in this way. In the rest of this section we will sketch the proof of Lemma 1, thereby seeing why this is the case. Without loss of generality we can set $F$ to be a diagonal positive semidefinite matrix with degenerate eigenvalues $\alpha_{i}$, each of degeneracy $d_{i}$. In particular let us partition $F$ into blocks determined by the degeneracies in its eigenvalues, i.e.
\begin{equation}
F=\left(
\begin{array}[c]{ccccc}
    F_{1} & 0 & 0 & 0 & 0\\
    0 & F_{2} & 0 & 0 & 0\\
    0 & 0 & . & . & . \\
    0 & 0 & . & . &  \\
    0 & 0 & . &   & . \\
\end{array}
\right),   \label{sigABCDEF}
\end{equation}
where the $F_{i}=\alpha_{i} I_{d_{i}}$, and $I_{d_{i}}$ is a $d_{i} \times d_{i}$ identity matrix. 

In \cite{Weig} it has been shown that if $F$ is of this form and both $FG^{\dag}$ and $G^{\dag}F$ are normal, then $G$ must have the same block structure as $F$, i.e.
\begin{equation}
G=\left(
\begin{array}[c]{ccccc}
    G_{1} & 0 & 0 & 0 & 0\\
    0 & G_{2} & 0 & 0 & 0\\
    0 & 0 & . & . & . \\
    0 & 0 & . & . &  \\
    0 & 0 & . &   & . \\
\end{array}
\right),
\end{equation}
and in addition we must have that each $G_{i}$ is a normal matrix of dimension $d_{i} \times d_{i}$. As each $G_{i}$ is normal, it can be diagonalised by a unitary transformation, say $U_{i}$. Hence the unitary matrix $W=\bigoplus_{i} U_{i}$ will diagonalise $G$ and leave $F$ unchanged (and hence diagonal). This corresponds to the physical operation $W \otimes W^{*}$. Hence any two matrices $F,G$ with the property that $FG^{\dag}$ and $G^{\dag}F$ are normal can be made Schmidt-correlated by local unitaries.The converse follows from performing a few lines of algebra to show that if $UFV^{T}=D_{1}$ and  $UGV^{T}=D_{2}$, where $D_{1}$ and $D_{2}$ are diagonal, then both $FG^{\dag}$ and $G^{\dag}F$ must be normal.

If there is no degeneracy amongst the Schmidt coefficients of $F$, then $G$ will be forced to have a diagonal block structure. Hence if $F$ has no degeneracy, $G$ {\it must} be diagonal if it is possible to express the states in Schmidt correlated form, and the states will be `obviously' Schmidt correlated from their Schmidt decomposition. If two pure states are not `obviously' Schmidt correlated in this way, it will only be possible to use local unitaries to make them so if there is sufficient degeneracy in the Schmidt coefficients of {\it both} states.

\section{Conclusion}

We have demonstrated that any two multipartite pure states can be inconclusively discriminated optimally 
using only local operations. We have also shown that this is possible for certain mixed states and 
certain regimes of conclusive discrimination. We then turned our attention to finding sets of
entangled states that can be recreated locally, thus allowing any global discrimination figure of merit to be 
achieved locally. We find that this is true for the Schmidt correlated states, and as a consequence this is also for any two 
maximally entangled states.

It would be interesting to know if there are many other states
which can be locally recreated using other techniques. If this can be shown
to apply to any two pure states, then we would know that two pure
states can be distinguished optimally under {\it any} figure of
merit using only local operations.

\par {\em Acknowledgements.} We would like to thank Jon Walgate and Vlatko Vedral for some illuminating discussions. We would also like to thank Koenraad Audenaart for drawing the results of \cite{Horn} to our attention. SV would also like to thank Sougato Bose for interesting conversations on this issue in the past. This work was supported by the Leverhulme Trust, the European Union EQUIP IST-FET project, the UK Engineering and Physical Sciences Research Council, the European Science Foundation QIT programme and the Italian Ministero dell'Universit\`{a} e della Ricerca Scientifica e Tecnologica.


\begin{thebibliography}{99}
\bibitem{Gen} If Alice and Bob possess $d$-level particles each, with each Schmidt correlated
state having Schmidt basis states $\{|k\rangle^{A}|k\rangle^{B} \}$, then the generalisation of this protocol
is for Alice to measure in a basis of states $\{|\chi\rangle \}$, with each $|\chi\rangle$ having the
property that $|\chi\rangle = 1/\sqrt{d}~ \Sigma_{k} \exp(i\theta_{k}) |k\rangle$. Such bases exist, 
so this protocol is possible.
\end{thebibliography}
\end{document}